\documentclass[11pt]{article}
\usepackage{amssymb}
\usepackage{amsmath}
\usepackage{setspace}
\usepackage[left=2.5cm,right=2.5cm,top=3cm]{geometry} 
\newcommand{\fn}{\footnote}

\newcommand{\cref}[1]{eqn.~(\ref{#1})}
\newcommand{\sref}[1]{Section~\ref{#1}}
\newcommand{\aref}[1]{Appendix~\ref{#1}}

\begin{document}
\title{A generalized equation of motion for gravity theories with non-minimal kinetic scalar couplings}
\author{Saugata Chatterjee\fn{schatte8@asu.edu}\vspace{0.2in}\\
Department of Physics, Arizona State University, \\
Tempe, Arizona 85287, USA \vspace{0.1in} \\
}

\date{}

\maketitle

\begin{abstract}
\noindent
A general form for the equation of motion for higher-curvature gravity is obtained. The interesting feature of the analysis is that it can handle Lagrangians which contain non-minimal kinetic scalar couplings. Certain subtle features, which are absent for the Einstein-Hilbert term, arise in higher-curvature gravity. These are identified and an algorithmic prescription is presented for the evaluation of the generalized equation of motion.
\end{abstract}

\vspace{0.2in}

\noindent
\section{Introduction}\label{sec:intro-eom}

It is common in theories like Kaluza-Klein gravity \cite{Salam:1981xd,Hinterbichler:2013kwa}, higher-curvature gravity \cite{Parikh:2009js,MuellerHoissen:1989yv,Meissner:1996sa}, and higher derivative scalar field theories \cite{Horndeski:1974wa,Deffayet:2009wt} to have various forms of non-minimal scalar couplings to gravity. Sometimes scalars are even kinetically coupled, like $G^{ab} \partial_a \phi \partial_b \phi $, where $G_{ab}$ is the Einstein tensor  \cite{Deffayet:2009wt,Kobayashi:2004hq}. A case by case evaluation of the equations of motion of these theories is a time-consuming process. The existence of a generalized expression for the equations of motion, applicable to a wide variety of non-minimally coupled theories, will greatly simplify the process; the goal of this paper would be to supply such an expression \cref{eq:gen-gr-eom4}. Existing efforts \cite{Mukhopadhyay:2006vu} in that direction builds upon the vision of Wald et al. \cite{Iyer:1994ys,Jacobson:1993xs} using a compact representation of the physical quantities (viz. entropy) in terms of generalized quantities. 
For example, when dealing with the entropy functional for non-Einsteinian theories of gravity an elegant representation of the generalized entropy can be made possible by defining an object 
\begin{align}
   P_a^{\phantom{a}bcd}= \partial L / \partial R^a_{\phantom{a}bcd}
\end{align}
This quantity facilitates the writing of the Wald entropy \cite{Wald:1993nt} in a more compact manner. This approach is the  motivation for obtaining a generalized expression for the equations of motion of higher-curvature gravity. 

In the following sections we will first derive a generalized equation of motion assuming the symmetries of the Riemann curvature tensor on the quantity $ P_{abcd} $. However our preliminary attempt will be met with failure, as several subtle features emerge. It appears to be obvious to assume that since the quantity $ P_a^{\phantom{a}bcd}$ is obtained by differentiating the Lagrangian with respect to the Riemann tensor, it should inherit the symmetry properties of the Riemann tensor $R_{abcd}$ \cite{Mukhopadhyay:2006vu}. However the assumption about the symmetries of the quantity $ P_{abcd} $ only works for theories which are homogeneous functions of the curvature scalars. For cases of non-minimal scalar kinetic couplings to gravity and in presence of the  derivatives of the Riemann curvature in the Lagrangian the quantity $ P_a^{\phantom{a}bcd}$ does not inherit the symmetry properties of the Riemann curvature. Therefore the procedure needs to be revised, so as to accommodate the symmetries of the quantity $P_{abcd}$ (or lack thereof). For example, in galileon theories \cite{Nicolis:2008in,deRham:2010eu} we encounter certain Lagrangians of the type $G^{ab} \partial_a \phi \partial_b \phi $ \cite{Deffayet:2009wt,Kobayashi:2004hq,VanAcoleyen:2011mj} which cause this procedure to fail. Our task will be to extend the paradigm in order to provide a more rigorous treatment of the derivation of the generalized equations of motion for non-minimally coupled Lagrangians (except the ones which contain derivatives of the Riemann curvature). That is, we will assume an arbitrary Lagrangian of the form $L(g_{ab},R_{abcd},\phi,\partial \phi )$. Throughout this paper, when we say ``equation of motion" we will always mean the gravitational equation of motion; the scalar equation of motion is relatively straightforward to derive. 

First we will perform a derivation of the generalized equations of motion using the assumption that the Lagrangian is a function of the curvature scalars only. Then in the subsequent section we will extend it to include all forms of non-minimally coupled Lagrangians (but not ones which contain derivatives of the Riemann curvature). At the end we will verify our procedure by evaluating the equations of motion for a non-minimally kinetically-coupled Lagrangian using brute force variational principle method and comparing the resulting expression with the one obtained from our expression for the generalized equations of motion \cref{eq:gen-gr-eom4}.

\section[EOM for Lagrangians without Kinetic Non-Minimal Couplings]{Derivation of Generalized Equations of Motion for Lagrangians without Kinetic Non-Minimal Couplings}\label{sec:gen-gr-eom1}

In this section we will derive a generalized equation of motion for a Lagrangian which is a homogeneous function of curvature scalars. It can also include non-minimal scalar couplings as long as the scalar couplings are not kinetic in nature. The procedure involves varying the Lagrangian with respect to the metric and the Riemann tensor separately and then reducing the varied expression further by expanding the variation of the Riemann tensor again. So it is not a Palatini variation even though we are varying two independent quantities at first. We will assume that the Lagrangian is a scalar function of $g^{ab}$ and $ R^a_{\phantom{a}bcd}$ only, $L(g^{ab}, R^a_{\phantom{a}bcd})$. The variation of the Lagrangian is \cite{Mukhopadhyay:2006vu}
\begin{eqnarray}
\delta \left(L\sqrt{-g}\right) &=& \left(\frac{\partial L\sqrt{-g}}{\partial g^{ab}}\right)\, \delta g^{ab} + \left(\frac{\partial L \sqrt{-g}}{\partial R^a_{\phantom{a}bcd}}\right)\, \delta R^a_{\phantom{a}bcd} \nonumber \\
&=& \left(\frac{\partial L\sqrt{-g}}{\partial g^{ab}} \right)\,\delta g^{ab} + \sqrt{-g} P_a^{\phantom{a}bcd}\, \delta R^a_{\phantom{a}bcd} \label{eq:gen-gr-eom0}
\end{eqnarray}
The variation of the determinant of the metric is $\delta log(g) = \delta g_{ab} ~ g^{ab}$ which implies the following relation. 
\begin{eqnarray}
  \delta \sqrt{-g} = - \frac{1}{2} \sqrt{-g} g_{ab} \delta g^{ab}
\end{eqnarray}
To evaluate $\delta R^a_{\phantom{a}bcd}$ we use the Palatini identity and use a gauge choice of the Riemann normal coordinates i.e. the Christoffel  are all zero but their derivatives are not. This allows us to replace the partial derivatives with covariant derivatives. And since the variation of the Christoffel is a tensor quantity, the whole expression becomes gauge-invariant and we no longer have to worry about whether we are in the Riemann normal coordinates or not.
\begin{eqnarray}
\delta R^a_{\phantom{a}bcd} &=& \delta \left[ \partial_c \left( \Gamma^a_{db}\right) - \partial_d \left( \Gamma^a_{cb}\right) \right] = \nabla_c \left(\delta \Gamma^a_{db}\right) - \nabla_d \left(\delta \Gamma^a_{cb}\right) \nonumber \\
&=& \frac{1}{2} \nabla_c\left[ g^{ai}\left(-\nabla_i\delta g_{db} + \nabla_d\delta g_{bi} + \nabla_b\delta g_{di} \right) + \delta g^{ai} \Gamma_{idb} \right]\nonumber - \{\textrm{term with } c\leftrightarrow d\} \\
&=& \frac{g^{ai}}{2} \nabla_c\left[-\nabla_i\delta g_{db} + \nabla_d\delta g_{bi} + \nabla_b\delta g_{di}\right] - \{\textrm{term with } c \leftrightarrow d\}  \label{eq:Palatini-identity}
\end{eqnarray}
Using this expression the last term in \cref{eq:gen-gr-eom0} can be simplified as 
\begin{eqnarray}
  P_a^{\phantom{a}bcd}\, \delta R^a_{\phantom{a}bcd} = P^{ibcd} \nabla_c\left[ -\nabla_i\delta g_{db} + \nabla_d\delta g_{bi} + \nabla_b\delta g_{di} \right] 
\end{eqnarray} 
where we have used the anti-symmetry of c and d indices. But $ P^{ibcd} \nabla_c \nabla_d\delta g_{bi} = 0 $ because of anti-symmetry in the indices $i$ and $b$ of $ P^{ibcd} $. The rest of the terms $ -\nabla_i\delta g_{db} + \nabla_b\delta g_{di} $ are anti-symmetric in $i$ and $b$. They contribute a total of $ 2 \nabla_b\delta g_{di} $ and we obtain
\begin{eqnarray}
P_a^{\phantom{a}bcd}\delta R^a_{\phantom{a}bcd} &=& 2 P^{abcd} \nabla_c\nabla_b (\delta g_{ad}) \nonumber \\
&=& 2 \nabla_c \left[ P^{abcd} \nabla_b \delta g_{ad} \right] - 2 \nabla_b \left[ \nabla_c P^{abcd} \delta g_{ad} \right] + 2 \nabla_b\nabla_c P^{abcd} \delta g_{ad} \label{eq:Pabcd-delta-Rabcd}
\end{eqnarray}
Discarding the total derivatives, only the last term contributes to the equation of motion. However, it needs to be rearranged so that the variation is with respect to $\delta g^{ab}$. This is easily done with raising and lowering indices and keeping track of the indices. A useful identity which we will use often is one involving contraction of the metric with both components of a two--tensor.
\begin{align}
  A^{ab} \delta g_{ab} = A_{mn} (g^{am} g^{an} \delta g_{ab}  ) = A_{mn} (\delta  g^{mn} - \delta  g^{am} g^{bn} g_{ab} - g^{am} \delta g^{bn} g_{ab}  ) = -  A_{mn} \delta  g^{mn}   \label{eq:Aab-delta-gab}
\end{align}
This identity, when used with the above expression leads to, $ 2 \nabla_b\nabla_c P^{abcd} \delta g_{ad} =  2 \nabla^c \nabla^d P_{acbd} \delta g^{ab} $, where we have used both the above identity and the anti-symmetry property of $ P_{abcd}$.
Now going back to \cref{eq:gen-gr-eom0} and collecting terms proportional to $\delta g_{ab} $ we obtain a generalized equation of motion
\begin{eqnarray}
  \frac{\partial L}{\partial g^{ab}} - \frac{1}{2} g_{ab} L + 2 \nabla^c \nabla^d P_{acbd} = 8 \pi G T_{ab}\label{eq:gen-gr-eom}
\end{eqnarray}
Care needs to be taken while differentiating the Lagrangian with respect to the metric, because there may be contractions with $g_{ab}$. The $\delta g_{ab}$ term contributes a negative term to the equations of motion. As an example, let us consider $R_{abcd} R^{abcd}$. We need to rewrite this as $ R^a_{\phantom{a}bcd} R^p_{\phantom{a}qrs} g_{ap} g^{bq} g^{cr} g^{ds}$ because  the independent variable is the Riemann tensor with one up index and not the one with all indices down.  The variation of the three $g^{ab}$ and one $ g_{ab}$ will give a total of 2 $R_{abcd} R_e^{\phantom{a}bcd}$. The symmetry factor is the number of contravariant metric contractions minus the number of covariant metric contractions. As an application of this procedure, the equations of motion for the Gauss-Bonnet gravity is obtained in \aref{app:eom-gb} using the generalized expression, \cref{eq:gen-gr-eom}.

\subsection*{Limitation of this method}\label{sec:fail-eom1}

The method to derive a generalized equation of motion outlined above makes several assumptions. While the procedure still works for non-minimally coupled scalar fields as shown at the end of \aref{app:eom-gb}, it fails for Lagrangians with a non-minimal kinetic scalar coupling, for example with $L = G^{ab} \partial_a \phi \partial_b \phi $. The failure is a result of the assumption that the  quantity $P_{abcd}$ has the same symmetries as the Riemann curvature tensor, which is impossible to impose if the Lagrangian is not a homogeneous function of the curvature. This is because our variation was carried out using $ R^a_{\phantom{a}bcd}$ and not $R_{abcd}$. While the latter has all the symmetries of the Riemann curvature tensor the previous one does not have the anti-symmetry in the first two indices. If the Lagrangian is not a homogeneous function of the curvature then $P^a_{\phantom{a}bcd}$ will have some up index term which will not inherit the symmetries of the curvature tensor when lowered. 
To illustrate this with an example we will derive the equations of motion for the above Lagrangian,  $L = G^{ab} \partial_a \phi \partial_b \phi $, by brute force variational method and then compare it with the one obtained from our generalized equations of motion. The calculation  has been carried out in \aref{app:gab-eom}. The equation of motion is
\begin{align}
  &  (R_{ac}\partial_b \phi + R_{bc} \partial_a \phi)\partial^c \phi    - \frac{1}{2} R \partial_a \phi \partial_b \phi   - \frac{1}{2} (\partial \phi)^2 R_{ab} + \nabla_a \nabla^c \phi \nabla_b \nabla_c \phi - \Box \phi \nabla_a \nabla_b \phi + R_{acbd} \partial^c \phi \partial^d \phi    \nonumber \\
  & \qquad \qquad   + \frac{1}{2} g_{ab} \left(  (\Box \phi)^2 - \nabla_c \nabla_d \phi \nabla^c \nabla^d \phi   \right) - \frac{1}{2} g_{ab} \left( - \frac{1}{2} (\partial \phi )^2 R \right) - g_{ab} R_{cd} \partial^c \phi \partial^d \phi  = 8 \pi G T_{ab}   \label{eq:Gab-brute-force-eom-app}   
\end{align}
We shall not go through the steps but the reader can  easily verify that this is not the same equation of motion which is obtained when \cref{eq:gen-gr-eom} is used.

\section[EOM for Lagrangians with Kinetic Non-Minimal Couplings]{Derivation of Generalized Equations of Motion for Lagrangians with Kinetic Non-Minimal Couplings}\label{sec:gen-gr-eom2}

We wish to find a generalized equation of motion from a generalized variational principle which applies to more general kinds of Lagrangians, not just the ones which are homogeneous  in the curvature scalars\fn{The Lovelock invariants are examples of homogeneous Lagrangians.}. The Lagrangian which is a function of the metric and its derivatives $L(g,\partial g, \partial \partial g \hdots ; \phi, \partial \phi)$ can be written in a locally flat gauge $\partial g \sim 0 $ (Riemann normal coordinates), which keeps the analysis covariant and the physics invariant. We also assume that the Lagrangian has derivatives of the metric (and all other fields) upto second order only. Any higher derivatives of the metric (or any other field) will introduce ghosts in the theory. The Lagrangian $L(g, \partial \partial g; \phi, \partial \phi)$ can now be varied as $\delta L = \mathbf{H}  \delta g + \mathbf{P}  \delta (\partial \partial g)$. However, the double derivative of the metric only appears as the curvature scalars and their coupling to matter fields. So we can rewrite the variation as $\delta L = \mathbf{H} \delta g + \mathbf{P} \delta R$. Now the equation of motion depends on whether we vary with respect to the Riemann with all down indices or one up index. In the previous case the equation of motion is much more compact and $\mathbf{P}$ inherits all the symmetries of the Riemann while in the latter case the equation of motion is more complicated but no symmetries needs to be imposed on $\mathbf{P}$. We shall begin with the latter case first.

The variation procedure is similar to the one depicted in \sref{sec:gen-gr-eom1}. All equations from \cref{eq:gen-gr-eom0} to \cref{eq:Palatini-identity} are still true, only we no longer assume or impose the symmetries of the Riemann curvature tensor on $P_{abcd}$. There is no \cref{eq:Pabcd-delta-Rabcd} and the  equation of motion is,
\begin{eqnarray}
  && \frac{\partial L}{\partial g^{ab}} - \frac{1}{2} g_{ab} L + \frac{1}{4} \nabla^c \nabla^d \left[ (  P_{abcd} - P_{abdc} ) + (  P_{bacd} - P_{badc} ) + (  P_{cadb} - P_{cabd} )  \right. \nonumber \\
  && \quad \left. + (  P_{cbda} - P_{cbad} ) + (  P_{bcad} - P_{bcda} ) + (  P_{acbd} - P_{acdb} )  \right] = 8 \pi G T_{ab} \label{eq:gen-gr-eom1} 
\end{eqnarray}
So far we have varied the Riemann tensor and not the Ricci tensor. However, if the Lagrangian contains only the Ricci tensor and not the Riemann tensor, as in the Einstein-Hilbert Lagrangian, then it makes sense to vary with respect to the Ricci tensor. The question is whether a generalized equation of motion derived from variation of the Ricci tensor would lead to a completely different equation of motion or whether our equation of motion can be reduced to the one obtained by variation of the Ricci tensor. To answer this let us find a generalized expression for the equation of motion using the variation of the Ricci tensor. We write the variation of the Lagrangian as,
\begin{eqnarray}
\delta \left(L\sqrt{-g}\right) &=& \left(\frac{\partial L\sqrt{-g}}{\partial g^{ab}}\right)\, \delta g^{ab} + \left(\frac{\partial L \sqrt{-g}}{\partial R_{ab}}\right)\, \delta R_{ab} \nonumber \\
&=& \left(\frac{\partial L\sqrt{-g}}{\partial g^{ab}} \right)\,\delta g^{ab} + \sqrt{-g} P^{ab}\, \delta R_{ab} \label{eq:var2}
\end{eqnarray}
After discarding some total derivatives, the equation of motion is
\begin{align}
   \frac{\partial L}{\partial g^{ab}} - \frac{1}{2} g_{ab} L + \frac{1}{4} \left( \nabla^d \nabla_d ( P_{ab} + P_{ba} ) + 2 g_{ab} \nabla^c \nabla^d P_{cd}  - \nabla^d \nabla_b ( P_{ad} + P_{da} ) \right. & \nonumber \\
   \left. - \nabla^d \nabla_a ( P_{bd} + P_{db} ) \right) = 8 \pi G T_{ab} & \label{eq:gen-gr-eom2}
\end{align}
This is the equation the previous \cref{eq:gen-gr-eom1} should reduce to. The $ P_{abcd}$ is actually related to $ P_{ab}$ by $P_{abcd} = g_{ac} P_{bd}$. If we use this expression for $ P_{abcd}$ in \cref{eq:gen-gr-eom1} we will obtain \cref{eq:gen-gr-eom2} thereby proving that the generalized equation of motion \cref{eq:gen-gr-eom1} works for all types of higher-curvature Lagrangians (without scalar kinetic couplings). We can further simplify \cref{eq:gen-gr-eom1} to,
\begin{align}
    \frac{\partial L}{\partial g^{ab}} - \frac{1}{2} g_{ab} L + \frac{1}{2} \left(  R_a^{\phantom{a}kcd} P_{(kb)cd} + R_b^{\phantom{a}kcd}  P_{(ak)cd}  \right)  +  \frac{1}{2}\nabla^c \nabla^d \left(  P_{ca[db]}  +   P_{cb[da]} \right. & \nonumber \\
    \left.  +  P_{bc[ad]} + P_{ac[bd]}  \right) =  8 \pi G T_{ab} & \label{eq:gen-gr-eom3} 
\end{align}
The symmetrization symbols have symmetry factors in them $(ab) = \frac{1}{2} (ab+ba) ; [ab] = \frac{1}{2} (ab-ba)$. 

At the beginning of this section we had claimed that we will obtain the generalized equations of motion by varying with respect to the Riemann tensor with one up index, $ R^a_{\phantom{a}bcd}$ and then obtain it with by varying with respect to the Riemann tensor with all down indices, $R_{abcd}$. We did the first one and now we proceed to do the latter. We do a similar variation as \cref{eq:gen-gr-eom1} but with the Riemann tensor with all indices down. Here the anti-symmetry of the Riemann tensor in all of its indices are manifest. Hence when we write down the $ P_{abcd}$ , it has to be correctly anti-symmetrized so that its indices have the same symmetries as the Riemann tensor. We can do this since all  index lowered $P_{abcd}$ will inherit the symmetry properties of the Riemann tensor, as explained in the previous section. The variation of the Lagrangian is,
\begin{eqnarray}
\delta \left(L\sqrt{-g}\right) && = \left(\frac{\partial L\sqrt{-g}}{\partial g^{ab}}\right)\, \delta g^{ab} + \left(\frac{\partial L \sqrt{-g}}{\partial R_{abcd}}\right)\, \delta R_{abcd} \nonumber \\
&& = \left(\frac{\partial L\sqrt{-g}}{\partial g^{ab}} \right)\,\delta g^{ab} + \sqrt{-g} P^{abcd}\, \delta R_{abcd} \nonumber \\
&& = \sqrt{-g} \left( \left( \frac{\partial L}{\partial g^{ab}} - \frac{1}{2} g_{ab} L \right)\delta g^{ab} + P^{acde} R^b_{\phantom{a}cde} \delta g_{ab} +  P_a^{\phantom{a}bcd} \delta R^a_{\phantom{a}bcd} \right) \qquad  \label{eq:var3}
\end{eqnarray}
There is a subtlety here. Even though we have defined the generalized quantity $ P_{abcd} $ as a variation of the Lagrangian with respect to $R_{abcd}$, the actual variation of the action still has to be with respect to  the Riemann tensor with one up index, $ R^a_{\phantom{a}bcd}$. This has to do with the way the Riemann curvature tensor is defined. It is defined with respect to  the Christoffel's with one index up. This adds an extra term $\mathbf{P} \cdot \mathbf{R}$  term to the variation which will be crucial to the analysis, as explained at the end. 

We can now use the Palatini identity (\cref{eq:Palatini-identity}) to simplify the last term of this expression in a similar fashion to \sref{sec:gen-gr-eom1}.
Going back to \cref{eq:var3} and collecting terms proportional to $\delta g^{ab} $ we have the generalized equations of motion,
\begin{eqnarray}
   \frac{\partial L}{\partial g^{ab}} - \frac{1}{2} g_{ab} L - \frac{1}{2} \left( \frac{}{} P_a^{\phantom{a}cde} R_{bcde} +  P_b^{\phantom{a}cde} R_{acde} \right) + \nabla^c \nabla^d \left(  P_{acbd} +  P_{bcad}  \right)= 8 \pi G T_{ab} \qquad  \label{eq:gen-gr-eom4}
\end{eqnarray}
where the explicit symmetrization of the indices a \& b is done as the equations of motion are expected to be symmetric in those indices. The rule regarding the derivative with respect to the metric, mentioned below \cref{eq:gen-gr-eom}, still applies and the symmetry factor is again the number of contravariant metrics minus the number of covariant metrics.  

This equation of motion \cref{eq:gen-gr-eom4} differs from the one in \cref{eq:gen-gr-eom3} in many ways. In \cref{eq:gen-gr-eom3} the $P_{abcd}$ has no symmetries at all while in \cref{eq:gen-gr-eom4} it has the symmetries of the Riemann tensor. The appearance of the $\mathbf{P} \cdot \mathbf{R}$ terms in the equations of motion is very interesting. Even though it appears that the $\nabla \nabla \mathbf{P} $ term in the equations of motion accounts for the variation of the Ricci or the Riemann tensor, it is in fact not entirely true. A part of it is proportional to $\delta \mathbf{R} $ but there are extra terms which are cancelled by the   $\mathbf{P} \cdot \mathbf{R}$ term. But there's more. The  $\mathbf{P} \cdot \mathbf{R}$  will still have a remaining piece  that cancels the extraneous terms which arises due to writing the Lagrangian in terms of the Riemann tensor when it can be expressed in terms of the Ricci tensor only. For example, if the Ricci scalar is written as $R_{abcd} \frac{1}{2} \left( \frac{}{} g^{ac} g^{bd} - g^{ad} g^{bc} \right) $ instead of $R_{ab} g^{ab} $. This point becomes apparent while evaluating the terms in the generalized expression (see \aref{app:gen-eom-Rab} and \aref{app:gen-eom-Gab})

For Lagrangians with a scalar field pre-factor, a further useful formula for the equations of motion can be written down using $L = e^{\mu \phi} L^\prime$ and $P_{abcd} = e^{\mu \phi} P_{abcd}^\prime$,
\begin{align}
    & e^{\mu \phi} \left( \frac{\partial L^\prime}{\partial g^{ab}} - \frac{1}{2} g_{ab} L^\prime - P_{(a}^{\prime\phantom{a}cde} R_{b)cde}  + \nabla^c \nabla^d \left(  P_{acbd}^\prime +  P_{bcad}^\prime  \right) \right. \nonumber \\
    & \qquad \qquad \qquad  \qquad \qquad \left.  \frac{}{}+ 2  P_{acbd}^\prime \left( \mu \nabla^c \nabla^d \phi + \mu^2  \partial^c \phi \partial^d \phi \right) \right)  = 8 \pi G  T_{ab} \label{eq:gen-gr-eom5}
\end{align}
In conclusion, we should either use \cref{eq:gen-gr-eom4} with the symmetries of the Riemann tensor imposed on the quantity $ P_{abcd}$ or use \cref{eq:gen-gr-eom3} without imposing any symmetries on the quantity $ P_{abcd}$. In the latter case $ P_{abcd}$ is simply the quantity which is left after the Riemann tensor is removed from the Lagrangian.
\subsection*{Discussion}

We obtained a generalized equation of motion for higher-curvature gravity. This expression works for most types of higher derivative Lagrangians except for Lagrangians which contains derivatives of the curvature scalars, viz. $L = \nabla_c R_{ab} \nabla^c R^{ab}$. However, these Lagrangians are third-order in the derivatives of the metric and will contain ghosts and therefore are not realistic models of gravity. Our expression is the most generalized expression possible for any type of Lagrangian with non-minimal kinetic scalar coupling.

\appendix

\section{Equations of Motion for Gauss-Bonnet Gravity}\label{app:eom-gb}

The Lanczos-Lovelock Lagrangians are special in that they are higher-curvature Lagrangians which have second order equation of motion. The first-order Lovelock Lagrangian is the Einstein-Hilbert term which forms the basis of Einsteinian general relativity and the second-order term is the Gauss-Bonnet Lagrangian. The Gauss-Bonnet Lagrangian is a topological term in four dimensions but has dynamical equations in higher dimensions. It is defined as,
\begin{align}
L_m = \delta^{b_1 b_2 b_3 b_4 }_{a_1 a_2 a_3 a_4} R_{a_1 a_2}^{\phantom{b_1}\phantom{b_2}b_1 b_2} R_{a_3 a_4}^{\phantom{b_1}\phantom{b_2}b_3 b_4} = R^2 - 4 R_{ab} R^{ab} + R_{abcd} R^{abcd}
\end{align}
The $P_{abcd}$ is straightforward to calculate. We just remove one Riemann $P^{\phantom{b}\phantom{b}cd}_{ab} =  \delta^{ c d b_3 b_4  }_{a b a_3 a_4  } R_{a_3 a_4 }^{\phantom{b_1}\phantom{b_2}b_3 b_4} $. The generalized $\delta$ is the Kronecker delta. We will need to express everything in terms of $R^a_{bcd}$ in order to derive the equation of motion. Therefore, the Lagrangian could be rewritten as $L = P^{\phantom{b}\phantom{b}cd}_{ab} R_{\phantom{b}\phantom{b}cd}^{ab} = g_{pa} g^{qb} g^{rc} g^{sd} P^{p}_{\phantom{b}qrs} R^{a}_{\phantom{b}bcd}$. The one down and three up metrics contribute a factor of two. And $\nabla^a P_{abcd} = 0$ for all Lovelock theories. We can now use this in the master equation \cref{eq:gen-gr-eom},
\begin{align}
	 P^{pqr}_a  R_{bpqr} - \frac{1}{4} g_{ab}  L = 4 \pi G T_{ab}
\end{align}
And using the specific form of $P_{abcd}$ for the Gauss-Bonnet
\begin{align}
 P_{abcd} = \frac{1}{2} \left( R ( g_{a c} g_{b d} - g_{ad} g_{bc} ) - 2 ( R_{ac} g_{bd} - R_{ad} g_{bc} - R_{bc} g_{ad} + R_{bd} g_{ac} ) + 2 R_{abcd} \frac{}{}\right) \label{eq:gb-Pabcd}
\end{align}
we get the equation of motion for the Gauss-Bonnet Lagrangian. Explicitly,
\begin{align}
	R_a^{\phantom{b}cde} R_{bcde} - 2 R_{ac} R^c_b -2 R^{cd} R_{acbd} + R R_{ab} - \frac{1}{4} g_{ab} ( R^2 - 4 R_{ab} R^{a b} + R_{abcd} R^{abcd}  )  = 4 \pi G T_{ab} \label{eq:gb-eom1}
\end{align}
This is a well known result. The real power of this general procedure is apparent when there are some non-minimal scalar couplings to the Lagrangian. For example, if we want the equation of motion for a Lagrangian of the form $L = e^{\mu \phi} ( R^2 - 4 R_{ab} R^{ab} + R_{abcd} R^{abcd} ) $ we repeat the procedure with $ \nabla^c \nabla^d P_{acbd}$ with $P_{abcd} = e^{\mu \phi} P_{abcd}^{GB}$ where the $P_{abcd}^{GB}$ is the one from minimally coupled Gauss-Bonnet Lagrangian, \cref{eq:gb-Pabcd}. This $P_{abcd}$ no longer satisfies the Bianchi identity $\nabla^a P_{abcd} = 0$ because of the non-minimal coupling. The equation of motion can be easily calculated by evaluating
\begin{align}
   &e^{-\mu \phi} \nabla^c \nabla^d P_{acbd} =  P_{acbd}^{GB}  \left( \mu \nabla^c \nabla^d \phi + \mu^2 \nabla^c \phi \nabla^d \phi \right)  \nonumber \\
   & = \mu \left( \frac{1}{2} R ( g_{ab} \Box \phi - \nabla_a \nabla_b \phi ) - R_{ab} \Box \phi - g_{ab} R_{cd} \nabla^c \nabla^d \phi + R_{ad} \nabla_b \nabla^d \phi + R_{bd}\nabla_a \nabla^d  \phi  \frac{}{}+ R_{acbd} \nabla^c \nabla^d \phi \frac{}{}\right) \nonumber \\
   & + \mu^2 \left( \frac{1}{2} R ( g_{ab} (\partial \phi )^2 - \partial_a \phi \partial_b \phi  ) - R_{ab}   (\partial \phi )^2 - g_{ab} R_{cd} \partial^c \phi \partial^d \phi + R_{ad} \partial_b \phi \partial^d \phi + R_{bd} \partial_a \phi \partial^d \phi \frac{}{} \right. \nonumber \\
   & \qquad \qquad \qquad \left.  \frac{}{}  + R_{acbd} \partial^c \phi \partial^d \phi\right) \label{eq:Pabcd-gb-phi}
\end{align}
We have intentionally multiplied the derivative of $P_{abcd}$ with $e^{-\mu \phi}$ to get rid of that pre-factor. The equation of motion is now the same as \cref{eq:gb-eom1} but with a factor $e^{-\mu \phi} $ on the stress tensor and with \cref{eq:Pabcd-gb-phi} included on the LHS of \cref{eq:gb-eom1}. 
 
\section{Derivation of the Equations of Motion for  $G_{ab} \partial^a \phi \partial^b \phi $  by Brute Force}\label{app:gab-eom}

Doing a straightforward variation of just $G_{ab} \partial^a \phi \partial^b $ (without the square root of the metric) leads to,
\begin{align}
  &\delta R_{ab}  \partial^a \phi \partial^b \phi + R_{ac}  \partial^c \phi \partial_b \phi \delta g^{ab} + R_{bc}  \partial^c \phi \partial_a \phi \delta g^{ab}  - \frac{1}{2} R \partial_a \phi \partial_b \phi \delta g^{ab}  - \frac{1}{2} (\partial \phi)^2 \delta R_{ab} g^{ab}  - \frac{1}{2} (\partial \phi)^2 R_{ab}  \delta g^{ab}   \label{eq:brute-var}
\end{align}
We can easily add the $-1/2 g_{ab} L $ term to the equation of motion after evaluating this variation. Before proceeding with the variation to Ricci tensor terms we will need the form of their variation which is easily obtained by contracting \cref{eq:Palatini-identity},
\begin{eqnarray}
  \delta R_{ab} = \frac{1}{2} \left( - \nabla^d \nabla_d \delta g_{ab} + \nabla^d \nabla_a \delta g_{bd} + \nabla^d \nabla_b \delta g_{ad} - \nabla_a \nabla_b ( g^{cd} \delta g_{cd} ) \right)
\end{eqnarray} 
A contraction with the metric gives,
\begin{eqnarray}
  \delta R_{ab} g^ {ab} = \nabla^a \nabla^b \delta g_{ab} - \nabla^d \nabla_d  ( g^{cd} \delta g_{cd} ) 
\end{eqnarray}
Now we are equipped to handle the variation of the Ricci terms,
\begin{align}
  - \frac{1}{2} (\partial \phi)^2 (\delta R_{ab} g^{ab} ) &= \left( \nabla_b \nabla_a \nabla^c \phi \partial_c \phi + \nabla_a \nabla^c \phi \nabla_b \nabla_c \phi - g_{ab}  ( \nabla_d \nabla^d \nabla^c \phi \partial_c \phi \right. \nonumber \\
  & \qquad \qquad + \left. \nabla_c \nabla_d \phi \nabla^c \nabla^d \phi ) \right) \delta g^{ab} 
\end{align}	
where we have used \cref{eq:Aab-delta-gab} and discarded the total derivatives. The other term is longer. We will just quote the final result.
\begin{align}
  \delta R_{ab}  \partial^a \phi \partial^b \phi  &= \frac{1}{2} \left( - 2 \Box \phi \nabla_a \nabla_b \phi  + g_{ab}( (\Box \phi)^2 + \nabla_c \nabla_d \phi \nabla^c \nabla^d \phi  +  \partial_d \phi \nabla^d \nabla_c \nabla^c \phi \right.  \nonumber \\
  & \qquad \qquad  \left.  + \partial_d \phi \nabla_c \nabla^d \nabla^c \phi )  -2 \partial_d \phi \nabla^d \nabla_a \nabla_b \phi  \right) \delta g^{ab} \label{eq:brute-force-eom-1}
\end{align}
Adding these two variations together,
\begin{align}
  &- \frac{1}{2} (\partial \phi)^2 (\delta R_{ab} g^{ab} ) + \delta R_{ab}  \partial^a \phi \partial^b \phi \nonumber \\
  &  = \left(  \nabla_a \nabla^c \phi \nabla_b \nabla_c \phi - \Box \phi \nabla_a \nabla_b \phi + R_{acbd} \partial^c \phi \partial^d \phi + \frac{1}{2} g_{ab} \left(  (\Box \phi)^2 - \nabla_c \nabla_d \phi \nabla^c \nabla^d \phi   \right) \right. \nonumber \\
  & \qquad \qquad \qquad \qquad \qquad  \left.  - \frac{1}{2} g_{ab} R_{cd} \partial^c \phi \partial^d \phi \right) \delta  g^{ab} 
\end{align}
The full equation of motion for  $G_{ab} \partial^a \phi \partial^b \phi $ is obtained by substituting this last expression in the \cref{eq:brute-var} and also adding the $-1/2 g_{ab} L $ term from the square root of the metric,
\begin{align}
  &  (R_{ac}\partial_b \phi + R_{bc} \partial_a \phi)\partial^c \phi    - \frac{1}{2} R \partial_a \phi \partial_b \phi   - \frac{1}{2} (\partial \phi)^2 R_{ab} + \nabla_a \nabla^c \phi \nabla_b \nabla_c \phi - \Box \phi \nabla_a \nabla_b \phi + R_{acbd} \partial^c \phi \partial^d \phi    \nonumber \\
  & \qquad \qquad   + \frac{1}{2} g_{ab} \left(  (\Box \phi)^2 - \nabla_c \nabla_d \phi \nabla^c \nabla^d \phi   \right) - \frac{1}{2} g_{ab} \left( - \frac{1}{2} (\partial \phi )^2 R \right) - g_{ab} R_{cd} \partial^c \phi \partial^d \phi  = 8 \pi G T_{ab}  
\end{align}

Since all the work is already done we can use the brute-force method to obtain the equations of motion for the Lagrangian $L = R_{ab} \partial^a \phi  \partial^b \phi $. This Lagrangian does not possess the symmetry properties of the Lagrangian $L = G_{ab} \partial^a \phi  \partial^b \phi $, which is a conserved quantity coupled with kinetic terms. The equation of motion for just $R_{ab} \partial^a \phi \partial^b \phi $ is,
\begin{align}
   &(R_{ac}\partial_b \phi + R_{bc} \partial_a \phi)\partial^c \phi - \frac{1}{2} g_{ab} R_{cd} \partial^c \phi \partial^d \phi  - \Box \phi \nabla_a \nabla_b \phi - \partial_d \phi \nabla^d \nabla_a \nabla_b \phi \nonumber \\
   &  + \frac{1}{2}g_{ab}( (\Box \phi)^2 + \nabla_c \nabla_d \phi \nabla^c \nabla^d \phi +  \partial_d \phi \nabla^d \nabla_c \nabla^c \phi + \partial_d \phi \nabla_c \nabla^d \nabla^c \phi )  = 8 \pi G T_{ab} \label{eq:Rab-brute-force-eom}
\end{align}
\subsection[EOM for $R_{ab} \partial^a \phi \partial^b \phi $ using Generalized EOM]{Equations of Motion for $R_{ab} \partial^a \phi \partial^b \phi $ using Generalized Equations of Motion} \label{app:gen-eom-Rab}
We are going to find the generalized equation of motion for the Lagrangian $L = R_{ab} \partial^a \phi  \partial^b \phi $ using the formula \cref{eq:gen-gr-eom3} in \sref{sec:gen-gr-eom2}. The $P_{abcd}$ with the right symmetries is
\begin{eqnarray}
  P_{acbd} &=&  \frac{1}{4} \left(  g_{ab} \partial_c  \phi \partial_d \phi - g_{ad} \partial_b \phi \partial_c \phi + g_{cd} \partial_a \phi\partial_b \phi  - g_{bc}\partial_a \phi \partial_d \phi \right)
\end{eqnarray}
and the derivative of the Lagrangian with the metric gives, 
\begin{eqnarray}
   \frac{\partial L}{\partial g^{ab}} = ( R_{bd} \partial_a \phi +  R_{ad} \partial_b \phi )\partial^d \phi +  R_{acbd} \partial^c \phi \partial^d \phi
\end{eqnarray}
We need the quantity  $\mathbf{P} \cdot \mathbf{R}$ in the equation of motion
\begin{eqnarray}
  P_a^{\phantom{b}cde} R_{bcde} = \frac{1}{2} R_{acbd} \partial^c \phi \partial^d \phi + \frac{1}{2} R_{bd} \partial_a \phi \partial^d \phi 
\end{eqnarray}
and its $a \leftrightarrow b$ counterpart $ P_b^{cde} R_{acde}$ together with the derivatives
\begin{eqnarray}
    && \nabla^c \nabla^d P_{acbd} = \frac{1}{4} \left(  -2 \partial_d \phi \nabla^d \nabla_a \nabla_b \phi - 2 \Box \phi \nabla_a \nabla_b \phi + g_{ab}( (\Box \phi)^2 + \nabla_c \nabla_d \phi \nabla^c \nabla^d \phi \right. \nonumber \\
  && \quad \qquad  \left.  +  \partial_d \phi \nabla^d \nabla_c \nabla^c \phi + \partial_d \phi \nabla_c \nabla^d \nabla^c \phi ) \right) + \frac{1}{4} R_{bd} \partial_a \phi \partial^d \phi - \frac{1}{4}   R_{acbd} \partial^c \phi \partial^d \phi 
\end{eqnarray}
and $\nabla^c \nabla^d P_{bcad}$ all added together in \cref{eq:gen-gr-eom3} leads to 
\begin{align}
   &(R_{ac}\partial_b \phi + R_{bc} \partial_a \phi)\partial^c \phi - \frac{1}{2} g_{ab} R_{cd} \partial^c \phi \partial^d \phi  - \Box \phi \nabla_a \nabla_b \phi - \partial_d \phi \nabla^d \nabla_a \nabla_b \phi \nonumber \\
   & \qquad \qquad +  \frac{1}{2}g_{ab}( (\Box \phi)^2 + \nabla_c \nabla_d \phi \nabla^c \nabla^d \phi +  \partial_d \phi \nabla^d \nabla_c \nabla^c \phi + \partial_d \phi \nabla_c \nabla^d \nabla^c \phi )  = 8 \pi G T_{ab} 
\end{align}
which is same as \cref{eq:Rab-brute-force-eom}.
\subsection[EOM for $G_{ab} \partial^a \phi \partial^b \phi $  using Generalized EOM]{Equations of Motion for $G_{ab} \partial^a \phi \partial^b \phi $  using Generalized Equations of Motion} \label{app:gen-eom-Gab}
In this section we see an explicit demonstration of the generalized method outlined in  \sref{sec:gen-gr-eom2} for the Lagrangian $L = G_{ab} \partial^a \phi  \partial^b \phi $. The $P_{abcd}$ with the right symmetries is
\begin{eqnarray}
  P_{acbd} &=&  \frac{1}{4} g_{ab} \left( \partial_c  \phi \partial_d \phi - \frac{1}{2} (\partial \phi)^2 g_{cd} \right) - \frac{1}{4} g_{ad} \left( \partial_b \phi \partial_c \phi - \frac{1}{2} (\partial \phi)^2 g_{bc} \right) \nonumber \\
  && + \frac{1}{4} g_{cd} \left(  \partial_a \phi\partial_b \phi - \frac{1}{2} (\partial \phi)^2 g_{ab} \right) - \frac{1}{4} g_{bc} \left( \partial_a \phi \partial_d \phi - \frac{1}{2} (\partial \phi)^2  g_{ad} \right)
\end{eqnarray}
and the derivative of the Lagrangian with the metric gives, 
\begin{eqnarray}
   \frac{\partial L}{\partial g^{ab}} = ( R_{bd} \partial_a \phi +  R_{ad} \partial_b \phi )\partial^d \phi +  R_{acbd} \partial^c \phi \partial^d \phi - \frac{1}{2} R \partial_a \phi \partial_b \phi - \frac{1}{2}  (\partial \phi)^2 R_{ab} \label{eq:Gab-eom1}
\end{eqnarray}
We are ignoring the term coming from the variation of the square root of the metric for the moment since it is trivially proportional to the Lagrangian itself. 
\begin{eqnarray}
  P_a^{\phantom{b}cde} R_{bcde} = \frac{1}{2} R_{acbd} \partial^c \phi \partial^d \phi + \frac{1}{2} R_{bd} \partial_a \phi \partial^d \phi - \frac{1}{2} R_{ab} (\partial \phi )^2 \label{eq:Gab-eom2}
\end{eqnarray}
The term left to calculate is the contribution by the variation of the Riemann,
\begin{align}
  \nabla^c \nabla^d P_{acbd} &=  \frac{1}{4} g_{ab} \left( (\Box \phi )^2 - \nabla_c \nabla_d \phi \nabla^c \nabla^d \phi \right) - \frac{1}{2} ( \Box \phi \nabla_a \nabla_b \phi - \nabla_a \nabla^d \phi \nabla_b \nabla_d \phi  )  \nonumber \\
  &   - \frac{1}{4} g_{ab} R_{cd} \partial^c \phi \partial^d \phi + \frac{1}{4} \left( R_{acbd}  \partial^c \phi \partial^d \phi + R_{bm} \partial^m \phi \partial_a \phi \right) \label{eq:Gab-eom3}
\end{align}
Adding up all these expressions (\ref{eq:Gab-eom1})  (\ref{eq:Gab-eom2}) (\ref{eq:Gab-eom3}) with their $a \leftrightarrow b  $ counterparts in \cref{eq:gen-gr-eom3} we get \cref{eq:Gab-brute-force-eom-app}.
\begin{align}
  &  (R_{ac}\partial_b \phi + R_{bc} \partial_a \phi)\partial^c \phi    - \frac{1}{2} R \partial_a \phi \partial_b \phi   - \frac{1}{2} (\partial \phi)^2 R_{ab} + \nabla_a \nabla^c \phi \nabla_b \nabla_c \phi - \Box \phi \nabla_a \nabla_b \phi + R_{acbd} \partial^c \phi \partial^d \phi    \nonumber \\
  & \qquad \qquad   + \frac{1}{2} g_{ab} \left(  (\Box \phi)^2 - \nabla_c \nabla_d \phi \nabla^c \nabla^d \phi   \right) - \frac{1}{2} g_{ab} \left( - \frac{1}{2} (\partial \phi )^2 R \right) - g_{ab} R_{cd} \partial^c \phi \partial^d \phi  = 8 \pi G T_{ab}    
\end{align}


\end{document}